# Simple and Accurate Computations of Solvatochromic Shifts in $\pi \rightarrow \pi^*$ Transitions of Aromatic Chromophores

by

Hendrik Heinz, Ulrich W. Suter,

Department of Materials, ETH, CH-8092 Zurich, Switzerland,

and

Epameinondas Leontidis,

Department of Chemistry, University of Cyprus, 1678 Nicosia, Cyprus

# Abstract

A new approach is introduced for calculating the spectral shifts of the most bathochromic $\pi \to \pi^*$ transition of an aromatic chromophore in apolar environments. As an example, perylene in solid and liquid *n*-alkane matrices was chosen, and all shifts are calculated relative to one well-defined solid-inclusion system. It is shown that a simple two-level treatment of the solute using Hückel theory yields spectral shifts in excellent agreement with experimental results for the most prominent inclusion sites of perylene in solid *n*-alkane surroundings and for the dilute solutions in liquid *n*-alkanes. The idea is general enough to be applied to any aromatic chromophore in a nonpolar solvent matrix. In contrast to earlier treatments, this approach is based on geometry-dependent polarizabilities, employs a $r^{-4}$ dependence for the dispersion energy, is conceptually simple and computationally efficient. Different simple models based on our general approach to compute the UV spectral shifts due to solvation indicate that the dispersive part of the van-der-Waals energy, which stabilizes the LUMO of perylene more than the HOMO, falls off with a distance dependence of $1/r^4$ in a range up to ~1 nm and not as $1/r^6$, as has been assumed for a long time. This finding corresponds to the interpretation of temporary dipoles as being equivalent to weak permanent dipoles with fluctuating orientation.

# 1. Introduction

A solvent-induced UV/Vis spectral red shift is due to the lowering of the transition energy between the ground and the excited state of a chromophore, caused by interactions of the chromophore-bearing (solute) molecule with the solvent molecules. We restrict ourselves to the case of $\pi \to \pi^*$ transitions of aromatic chromophores in nonpolar solvents, in which permanent electrostatic interactions and hydrogen bonding are absent[1] and where, therefore, the predominant interaction is dispersive. For the shifts in UV spectra of aromatic chromophores, environments of noble gases,[2-6] of alkanes,[7-10] and of polymeric nonpolar media[1] have been the primary targets in previous studies; several models have been proposed,[2-6, 11-15] and considerable experimental data, mostly on chromophores in alkane crystals, have been collected.[2-4, 16-24] Nevertheless, the theoretical concepts are far from simple and the concomitant computational approaches are difficult.

If we consider just one solvent unit, which may be a single atom in the simplest case, this unit is interacting with the dye molecule in its ground state and (after excitation) in its excited state. The only difference between the two states is the promotion of one electron from the HOMO into the LUMO. Since we consider a single electron to be promoted, the only possible cause for the spectral red shift must be a solvent-induced change in the dispersive interaction of this electron with the solvent unit that cannot be the same for the HOMO and the LUMO, since otherwise they would cancel. Since the direction of the spectral shift is always towards lower energies and dispersive energies are always stabilizing, we conclude that the electron in its excited state (LUMO) has a stronger dispersive interaction with the solvent unit than in the ground state (HOMO). It is often assumed that this is due to a higher diffusivity of the electron density within the LUMO; this idea is at the heart of the present treatment and is represented in Figure 1. The spectral shift, $\delta E$, is the difference of the dispersive energy between the solvent and the dye molecule with one electron promoted into the LUMO, $\delta E_{i0}$, and with the electron in the HOMO, $\delta E_{00}$:

$$\delta E = \delta E_{i0} - \delta E_{00}. \qquad (4)$$

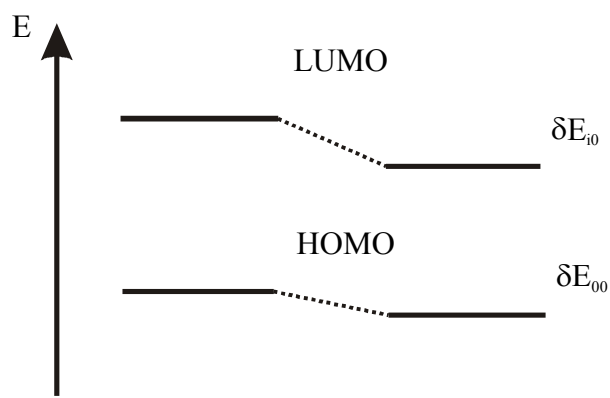

**Figure 1.** The relative energies of HOMO and LUMO of the solute molecule in vacuum and in a solvent matrix. Solvation lowers the energy through dispersive interactions, which are greater for the LUMO.

Here, our goal is to develop a quantitative understanding of environmental effects on these energies and thereby to obtain a simple, yet accurate method of estimating the most bathochromic $\pi \to \pi^*$ transition of an aromatic chromophore in apolar surrounding.

## 2. Critical Review on Calculations of Dispersive Interactions with Perturbation Theory

The earliest approach for the calculation of UV/Vis spectral shifts is that of Longuet-Higgins and Pople,[12] who used time-independent perturbation theory for the description of dispersive interactions. Shalev et. al.[3] extended this work. The dispersive energy arises from fluctuating dipole moments of the otherwise nonpolar molecules. This temporary dipole-dipole interaction energy is contained in the perturbation Hamiltonian, $H'$. Assuming point-like atom-based dipoles, their pair-wise interaction energy varies like $r^{-3}$, where $r$ is the distance between the centers of the two dipoles. In fact, this interaction energy depends on the orientation of the two dipoles at small and medium distances and converges to an $r^{-6}$ distance dependence when the distance is ~6 times the dipole length or more.[25, 26] In either case the first order correction to the energy,

$$E^{(1)} = \langle A_0 B_0 | H' | A_0 B_0 \rangle ,  \qquad (1)$$

is zero because the expectation values of the dipole moment operators in $H'$ are zero for non-polar atoms in their ground state. The second order correction to the energy, when both atoms are in the ground state, is given by

$$E^{(2)} = \sum_{j>0}\sum_{k>0} \frac{\langle A_0 B_0|H'|A_j B_k\rangle\langle A_j B_k|H'|A_0 B_0\rangle}{-(E_j + F_k)} \tag{2}$$

where $E_j$ and $F_k$ are the energies of the atoms $A$ and $B$ in their $j$-th and $k$-th state. The dipolar interaction was usually approximated as an $r^{-3}$ distance dependence,[3, 12] and the closure approximation[27] for the excited-state energies of atoms $A$ and $B$ yields directly London's formula for dispersive interactions (see ref. 27 for details):

$$E^{(2)} \propto \frac{\alpha_A \alpha_B}{r^6}. \tag{3}$$

However, as we pointed out, the dipole-dipole energy falls off as $r^{-6}$ at long ranges[26] so that perturbation theory actually yields a strange $r^{-12}$ distance dependence by the same procedure. Therefore, perturbation theory might not be a suitable approach to calculate such energies. Eq 3 or similar expressions relate to the dispersive energy between two atoms. The same approach can be used to obtain the dispersive energy of molecules, but the entire chromophore has to be considered instead of atom $A$ and a solvent unit in the place of atom $B$. For the chromophore, the total wave function is split into a linear combination of all atomic orbitals although this disconnects the included atoms. For the solvent, quasi-spherical entities with an "atomic" wave function are considered, e. g., a methyl group or a methylene group. The spectral shift from eq 1 with the two dispersion energies from the excited state $i$ and the ground state $0$ of the solute molecule $A$, assuming the perturbation approach, is:

$$\delta E_B = (\delta E_{i0} - \delta E_{00})_B = \sum_{j \neq i}\sum_{k>0} \frac{\langle A_i B_0|H'|A_j B_k\rangle^2}{E_i-(E_j+F_k)} - \sum_{j>0}\sum_{k>0}\frac{\langle A_0 B_0|H'|A_j B_k\rangle^2}{-(E_j+F_k)} \tag{4}$$

This is the spectral shift caused by one solvent unit $B$; to obtain the total spectral shift, one must add the contributions of all solvent units.

If eq 4 is implemented, the following needs to be considered: (1) The distance dependence for intermediate-range intermolecular interactions is better approximated by an $r^{-4}$ dependence than by one that goes as $r^{-6}$ (London's formula). Also, the obtained $r^{-6}$ distance dependence with perturbation theory is quite arbitrary, as discussed above. Some theories are based on the $r^{-6}$ dependence for long-range dispersive interactions[2, 4, 5, 10, 12] but are, generally speaking, not very accurate. An $r^{-4}$ dependence has been shown several times to better represent the intermediate range interactions.[1, 6, 8, 28] In what follows, we will also assume that an $r^{-4}$ dependence is better (trials of the methods described in section 5 with an $r^{-6}$ dependence yield less satisfactory results). (2) By inserting the LCAO expansion of a complex chromophore[3, 6] into the perturbation expression eq 4, one explicitly considers only two-atom interactions (the treatment was initially formulated for the interaction of two atoms). However, an aromatic chromophore has delocalized π electrons, the polarizability of which is strongly anisotropic and dependent on the environment. The delocalization of the electrons between the atoms or the existence of nodal planes are not considered. (3) The evaluation of eq 4 then yields $n^4$ terms in the summation,[3] when $n$ is the number of terms in the LCAO expansion — all the two-body interactions, which incorporate one solute atom and the solvent unit, and, in addition, a large number of three-body interactions,[3, 6] which incorporate all possible pairs of atoms in the solute molecule plus the solvent unit. The two-body terms are proportional to the square of the orbital coefficient on one solute atom, i.e., to the local electron density. The three-body terms have no explicit physical meaning because they contain products of orbital coefficients at two solute atoms in all combinations, each of them belonging to a different state, and somewhat arbitrary geometry factors.[3] These three-body terms constitute an essential part in this approach, even though a physical justification has never been given.[1, 3, 6, 8] (4) The evaluation of the matrix elements in eq 4 with appropiate electrostatic potentials is almost impossible, especially considering the numerous two-body and three-body terms arising out of the LCAO expansion. A simple two-atom interaction with $n^1$ terms instead of $n^4$ (see Section 3.5.2) gives results of the same precision. (5) From the computational point of view, the perturbation approach is not efficient. For a chromophore such as perylene, for instance, 92 states are included in the LCAO expansion and all of them are, in principle, used in the calculations. Overall, more than $10^6$ terms in the perturbation summation eq 4 must then be computed per solvent unit[3, 6] whereas our method featured below requires only 40 terms per solvent unit and is more accurate (see Table 4).

In conclusion, the approach of Shalev and Jortner et. al.[3] can hardly be recommended because an excessive number of matrix elements in eq 4 is required, which are of uncertain physical meaning and evaluated with several arbitrary scalings, as well as an $r^{-6}$ distance dependence of the dispersive energy (see the details in ref. 3). The underlying concept in perturbation theory that the (dispersive) interaction energy is summed over any possible combinations of states of each of the two atoms is only a postulate. We evaluate the dispersive energies in eq 4 in a direct way (see below): using an $r^{-4}$ distance dependence of the dispersive energy and employing explicit polarizabilities at the locations of measurable electron density, taking into account the shape of the electron clouds (which is not the case with perturbation approaches). A justification may be as follows. As a direct consequence of the Coulomb-law, we obtain an $r^{-3}$ distance dependence of the energy between two permanent dipoles for short ranges (distance ~ dipole length) and an $r^{-6}$ distance dependence between two permanent dipoles at long distances (distance > 6 × dipole length).[26] If, therefore, our two atoms or electron clouds can be understood as two weak permanent dipoles, the interaction energy in a medium distance range might be proportional to $r^{-4}$ so that $E \propto \alpha_A \alpha_B / r^4$.

Some other methods are based on "single-center" molecular polarizabilities,[2, 4] which are even simpler and computationally less demanding. With such an approach, Adams and Stratt[4, 5] have obtained good results for benzene approximated as a spherically-symmetric solute. However, this approach cannot be generalized for molecules of less symmetry, and the precision becomes considerably smaller.

## 3. Systems Considered and Methods Employed

### 3.1. The Chromophore-alkane System

The systems considered here are (1) inclusion sites of perylene in solid *n*-alkanes, and (2) dilute solutions of perylene in liquid *n*-alkanes.

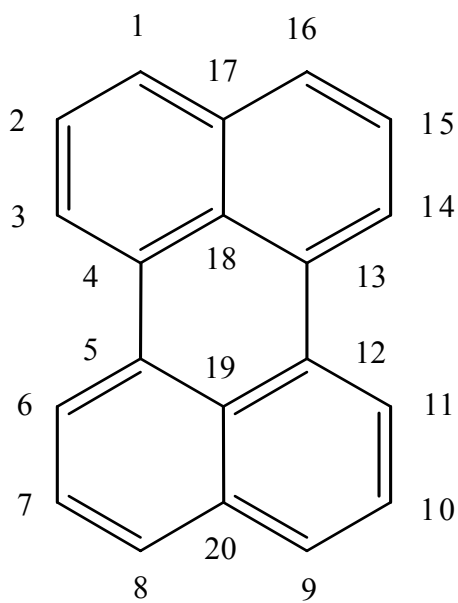

**Figure 2.** The numbering system employed for perylene and its orbital coefficients for HOMO and LUMO. The absolute values of the orbital coefficients at the atoms 1 to 4 are denoted $c_1$ to $c_4$, at all other atoms they are obtained by symmetry. Their relative signs can be derived from the MO pictures in Figure 3. The LUMO coefficients are multiplied with a factor $x$ to account for the increased polarizability of the electrons in the excited state (see text).

(1) In solid *n*-hexane, a perylene molecule (see Figure 2) can be inserted into the lattice by replacing two alkane molecules so that the molecular plane of perylene is either parallel to the *bc* or to the *ab* plane of the crystal axis system.[1] If it is parallel to the *bc* plane, the long axis of perylene has two possible orientations: it can be parallel to the long axis (*b* axis) of the lattice, corresponding to inclusion "site 1", or rotated against it by 60°, corresponding to the inclusion "site 2". If it is parallel to the *ab* plane, the long axis of perylene is parallel to the *b* axis, yielding inclusion "site 3".[1] These sites are very well defined in the total luminescence spectra (TLS, see Section 3.3). Leontidis et al.[9] have shown that for higher *n*-alkanes, inclusion site 1 exists and corresponds to a sharp, intense peak in the TL spectrum, but that there are no clearly identified arrangements corresponding to sites 2 and 3; instead, we find an ill-defined inclusion site with a considerably lower spectral shift and only a weak structural analogy to site 3. We employ the well-defined sites, the three sites in *n*-hexane and sites 1 of *n*-heptane to *n*-nonane to test the model developed.

(2) In solution, the molecular environment of any molecule is relatively vaguely defined. We consider the spectral shift of perylene in dilute solutions of *n*-alkanes with 6 to 10 carbon atoms, estimated from a number of thermalized snap-shots for every system and compared to experimental values (see Section 3.3).

### 3.2. Molecular Modeling

The Shpol'skii (solid inclusion) systems of perylene in *n*-alkanes were modeled starting with crystals defined by periodic boxes of 300 to 400 alkane molecules; room for a single perylene inclusion was obtained by replacing two or three of the alkane molecules, depending on the specific inclusion site studied.[9] All molecular simulations were carried out with the Discover program[8] and the Insight 400 graphic interface. The energy of the generated structures were first minimized by molecular mechanics, then the solid was equilibrated by NVT molecular dynamics for about 40 ps (time step 1 fs, Verlet's integrator, temperature control by velocity scaling). During the next 50 ps, 100 snapshots were collected and for each of them, the spectral shift for perylene estimated. The final values are the averages of those spectral shifts.

The dilute solution systems were constructed by placing 60 *n*-alkane molecules and one perylene molecule in a cubic cell with the overall density of the liquid n-alkane at 300 K and 1 bar. After energy minimization, the system was exposed to NVT and NpT molecular dynamics for 600 to 1000 ps, and during the next 200 ps, 100 snapshots are taken to calculate the spectral shifts (for the NpT simulations, Andersen's manostat was used with a cell mass of 100 and a pressure of 0.16 GPa to maintain the density). During the sampling interval, the molecules on average diffuse about three to six times the box edge length; the calculated shifts for each snapshot are not noticeably correlated. The spectral shifts obtained through NpT and NVT simulation are not significantly different.

### 3.3. Experimental

The spectral shifts for the solid inclusion sites, measured with total luminescence spectroscopy (TLS), have all been taken from the literature.[8, 19, 23, 24, 29] The spectral shifts in the dilute alkane solutions were measured using a LAMBDA 9 Perkin-Elmer UV/VIS/NIR spectrometer with a slit width of 0.5 nm. The solutions had a concentration of about $1.5 \cdot 10^{-4}$ M. At this concentration we assume that $\pi$-stacking interactions due to mutual instantaneous polarization of perylene molecules do not frequently occur.[14] The results are included in Table 4.

## 4. The Electronic States

### 4.1. Aromatic Solute

We employ a standard semi-empirical method, appropriate for the chromophore in question (perylene), such as the Hückel method, Extended Hückel method, MNDO, or AM1.[30] From these we obtain for each electronic state its symmetry, its energy, and its orbital coefficients (the square of which represents the electron density at each atom). It has to be decided, which electrons exert influence in the excitation process. The wave function of the solute molecule written as a linear combination of atomic orbitals is

$$\Psi_{total} = \sum_i c_i \phi_\sigma + \sum_j c_j \phi_\pi = \Psi_\sigma + \Psi_\pi \tag{5}$$

where the $\phi_\sigma$ are the atomic orbitals of the $s$ electrons and the $\phi_\pi$ are those of the $\pi$ electrons. All semi-empirical methods surmise that for any of the states, the contribution of the $s$ electrons, $\Psi_\sigma$, is orthogonal to that of the $\pi$ electrons, $\Psi_\pi$ ($\langle \Psi_\sigma | \Psi_\pi \rangle = 0$). Thus, there will be no changes in the distribution of the $s$ electrons during a $\pi \rightarrow \pi^*$ transition and the considerations can be limited to $\pi$ electrons only. Consequently, it is also possible to ignore all hydrogen atoms on the aromatic molecule.

It is also unlikely that all $\pi$ states need to be considered, since we are interested in $\pi \rightarrow \pi^*$ transitions only. In order to assess the relevance of particular states, the energy spacings of the $\pi$ orbitals around the HOMO and LUMO need to be examined. For our example of perylene, the corresponding values are listed in Table 1. Here we assign the relative energy 1.0 to the frontier orbital transition. Table 1 shows that the energy spacing to the neighboring orbitals is of the same magnitude, which indicates that to a good approximation we can neglect electronic states other than the HOMO and LUMO for the transition of concern (interestingly, the different semi-empirical methods give rather different values!). By considering only the HOMO and LUMO, the problem is considerably reduced.

**Table 1:** The relative energy spacing near the frontier orbitals in perylene as given by various semi-empirical methods.

| π orbitals considered | Hückel | Extended Hückel | MNDO | AM1 |
|---|---|---|---|---|
| [HOMO - 1] → HOMO | 0.39 | 0.77 | 1.41 | 1.37 |
| HOMO → LUMO | 1.00 | 1.00 | 1.00 | 1.00 |
| LUMO → [LUMO + 1] | 0.39 | 1.08 | 1.29 | 1.25 |

The frontier orbitals of perylene are graphically presented in Figure 3. There are nodal planes along all the principal molecular axes in the HOMO and along two of the principal axes in the LUMO. Only four different absolute values for the orbital coefficients exist in both states considered: these values, $c_1$ to $c_4$, are assigned to carbon atoms 1 to 4 in Figure 3. Their signs can be taken from the symmetries in Figure 3. In the HOMO, we can distinguish between two types of electron clouds, each spread over two atoms and occurring four times. In the LUMO, there are three types of electron clouds, two spread over two atoms and one localized on one atom only (see Figure 3).

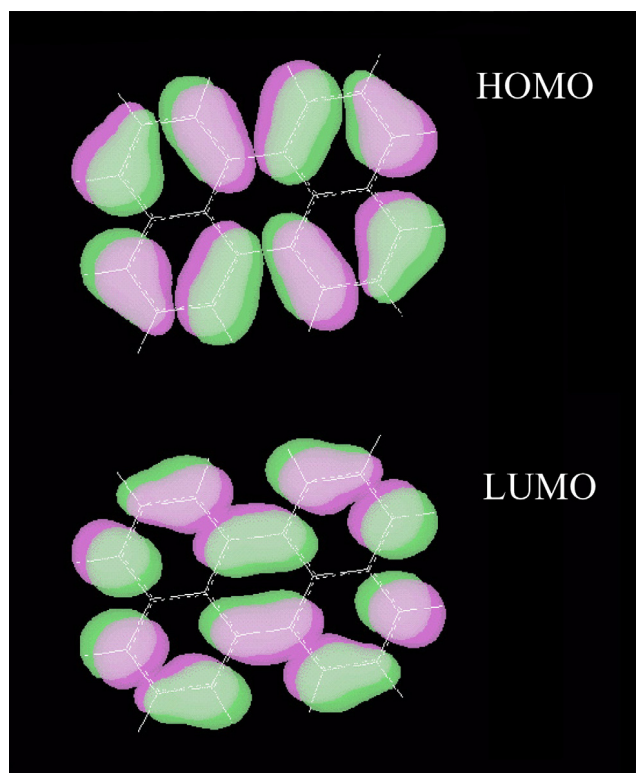

**Figure 3.** HOMO (above) and LUMO (below) of perylene. There are two nodal planes along the long axis of the molecule. In the HOMO, two differently shaped lobes of electron density can be found that are spread over two atoms. In the LUMO, three different lobes of electron clouds are observed; one of them is localized on one atom only. All these lobes occur repeatedly for symmetry reasons.

Since the semi-empirical methods do not all give the same results, one must be chosen for a particular piece of work. The four principal orbital coefficients are compared in Table 2. The differences between them are not significant, and we found that the particular choice of method does not change the results of the shift calculations significantly. The Hückel method was chosen for its simplicity and because it has no adjustable parameters.

Table 2: Comparison of the absolute values of the orbital coefficients generated by several semi-empirical quantum mechanical methods.

| coefficient | Hückel | Extended Hückel[a] | MNDO | AM1 |
|---|---|---|---|---|
| $c_1$ | 0.3283 | 0.3267 | 0.3154 | 0.3186 |
| $c_2$ | 0.1140 | 0.1085 | 0.1003 | 0.1007 |
| $c_3$ | 0.2887 | 0.2972 | 0.2957 | 0.2943 |
| $c_4$ | 0.2143 | 0.2078 | 0.2302 | 0.2274 |

[a] In the ground state.

Consider the difference in the electron distribution between the HOMO and LUMO. If the electron densities, represented by the square of the orbital coefficients, are the same at every atom in both states, the polarizabilities should be similar because the polarizability is dependent on electron (de)localization. As a result, only a minimal spectral shift would be observed. Because there are more nodal planes in the LUMO, which tends to increase electron localization, we might even expect the polarizability to diminish upon excitation and the dispersive stabilization to be less than in the HOMO, resulting in a positive spectral shift. However, experience shows that the reverse is true. The cause is probably that LUMO electrons posses a higher energy and are more delocalized than those in the HOMO. This electron mobility must be perpendicular to the plane of the aromatic molecule, because horizontally there are more restrictions in space caused by the extra nodal planes. Thus the anisotropy of the polarizability increases in the excited state.[31] Excited state polarizabilities are not known locally in larger molecules, and only roughly yet for some complete simple molecules.[31-35] The simplest way to account for the "diffusivity" of the electrons in the LUMO is to relay on the additional nodal planes. We propose to include the cross terms $c_i c_j$ in the normalization of the orbital coefficients,

$$\sum_i c_i^2 + \sum_{\substack{\text{connected atoms} \\ i \text{ and } j}} c_i c_j = 1 \qquad (6)$$

which will produce negative contributions for nodal areas between connected atoms, thereby increasing the orbital coefficients. The scaling factor $x$ for the orbital coefficients derived from eq 6 is:

$$x = \sqrt{1 - \sum_{\substack{\text{connected atoms} \\ i \text{ and } j}} c_i c_j} \ . \tag{7}$$

This factor will be used to scale the orbital coefficients of the LUMO relative to the HOMO ($x_{\text{LUMO}}/x_{\text{HOMO}}$); the coefficients for the HOMO are normalized in the usual way, without cross terms, and represent the real electron densities. For perylene, we obtain $x_{\text{LUMO}}/x_{\text{HOMO}} = 1.192$, i.e., in the LUMO the electrons are roughly 42 % more polarizable ($1.192^2 \approx 1.42$). It is interesting that even deviations in this ratio from 1.15 to 1.25 do not alter the results in a significant manner. The experimentally determined polarizability of the whole perylene molecule in the excited state is indeed 40 % higher than in the ground state.[36]

To sum up, we consider only the $\pi$ electrons of the chromophore in the frontier orbitals, using the simple Hückel-MO scheme. The increased "mobility" of the electrons in the LUMO is estimated with the aid of cross terms including products of the orbital coefficients.

### 4.2. Solvent

The alkanes in our case are divided into methyl and methylene groups and are treated as quasi-atoms. This approach works well because the electrons that give rise to dispersive forces are localized on these groups. Methyl and methylene groups have different polarizabilities (2.22 and 1.84 Å$^3$).[37, 38] The effect of a solvent group on the solute molecule transition is then calculated, and the contributions of each group are independently summed.

## 5. Computations of Spectral Shifts

For our model case, perylene, the symmetry of the electron distribution in the HOMO and in the LUMO are first calculated plus the numerical values for the four distinct orbital coefficients. Bond polarizabilities are then taken from the literature to describe the direction-dependent polarizabilities of the various electron clouds in the frontier orbitals. This and the geometry of the system allow then to compute the dispersive interaction between the electron clouds and the solvent units. This procedure is executed for all relevant electron clouds in the HOMO and LUMO, yielding the dispersion energies in both electronic states. The desired red shift is finally calculated using eq 1.

The dispersion energy between one solvent molecule $B$ and the solute molecule $A$ is given by

$$\delta E_D = -k\alpha_B \sum_i \frac{\alpha_i}{r_{A,B}^4}, \qquad (8)$$

where the summation over $i$ comprises all carbon atoms of $A$ and $\alpha_i$ is the polarizability at atom $i$. The distance dependence for medium-range dispersion forces was discussed in section 2; all alkyl groups with a distance of more than 1000 pm will be omitted since their contribution to the shift is not significant. The value of $\alpha_i$ is calculated using bond polarizabilities from the literature,[37, 38] see below. $\alpha_i$ depends on the position of the solvent unit relative to the respective electron cloud on the solute. $\alpha_B$ is the polarizability of the alkyl group considered, and $k$ is a fitting constant used to adjust the spectral shift of one selected line; for perylene, inclusion site 1 in $n$-hexane[8] was used (see Section 3.1). The total spectral shift for solvent unit $B$ is then given by

$$\delta E_B = (\delta E_{i0} - \delta E_{00})_B = -k\alpha_B \sum_i \frac{\alpha_i^{excited} - \alpha_i^{ground}}{r_{A,B}^4}, \qquad (9)$$

where $\alpha_i^{excited}$ and $\alpha_i^{ground}$ are the polarizabilities of the electron cloud at carbon atom $i$ for the excited and the ground state, respectively. This is the spectral shift caused by one solvent unit $B$; to obtain the total spectral shift, one must add the contributions of all solvent units.

### 5.1. Method A: The General Approach

The polarizabilities $\alpha_i^{excited}$ and $\alpha_i^{ground}$ are obtained as follows. The shape of the electron clouds is between that of a single $\pi$ bond and that of a singly occupied $p$ orbital (see Figure 3). The polarizability tensors for the C=C and the C-C bond are known[38] and that of a single $p$ orbital can be deduced from the polarizability of a methyl radical[37] minus the bond polarizabilities for the C-H bonds.[38] For the $\pi$ bond, we estimate the polarizabilities in the direction of the principal axes as the difference between those of a double bond and a single bond (see Table 3); the bond lengths of the two bonds are different (133 vs. 154 pm) and also different from that of an aromatic bond (141 pm), but calculations with slightly changed values give essentially identical results (bear in mind that the electron localization description on the basis of the shape of an ordinary $\pi$ bond is already a considerable approximation). The similarity in the values for $\alpha_{yy}$ and $\alpha_{zz}$ (see Table 3) suggests cylindrically symmetric aromatic bonds and we surmise that $\alpha_\parallel^0 = 1.83$ Å$^3$ and $\alpha_\perp^0 = 0.49$ Å$^3$. (Here we use the conventional unit Å: 1 Å$^3$ = 10$^{-30}$ m$^3$) For the single *p* orbital we assume an isotropic polarizability and estimate it from that of a methyl radical, calculated using Miller's[13] atomic polarizability increments to 3.088 Å$^3$. After subtraction of the polarizability of three C-H bonds (3 · 0.65 Å$^3$) one obtains $\alpha_p$ = 1.14 Å$^3$. Modification of this value between the limits $\alpha_p$ = 0.88 and 1.23 Å$^3$ still gave excellent agreement with experiment. The best value overall seems to be $\alpha_p^0$ = 1.06 Å$^3$.

**Table 3:** Polarizability components for various occupied lobes, from Ref. 38, in Å$^3$. The *x* direction is parallel to the given bond, *y* and *z* perpendicular so that the *z* axis is in the direction of the *p* orbitals forming the π bond.

| electron space | $\alpha_{xx}$ | $\alpha_{yy}$ | $\alpha_{zz}$ |
|---|---|---|---|
| C=C bond | 2.80 | 0.73 | 0.77 |
| C-C bond | 0.97 | 0.26 | 0.26 |
| π bond (estimated) | 1.83 | 0.47 | 0.51 |

One must now calculate the values for $\alpha_\parallel$ and $\alpha_\perp$ for each type of electron cloud in the HOMO and LUMO because the shape of these lobes deviates considerably from that of a "normal" π bond (see Figure 3). A normal π bond is centered at two carbon atoms and characterized by a cloud with an integral charge density of unity and the same electron density at both atoms, while a singly occupied π orbital has the electron density zero at one of the two atoms. A crude approximation might involve scaling the two atomic polarizability components linearly between these two extreme cases, using the ratio of the electron densities at the two carbon atoms as a scaling parameter. The physical justification rests on the fact that the polarizability in a given direction depends on the electron density; also, this approach avoids introducing new parameters. Hence, assuming that exactly one electron is localized in the cloud, we scale the longitudinal and vertical components of the polarizability of atom *i* as shown in Figure 4.

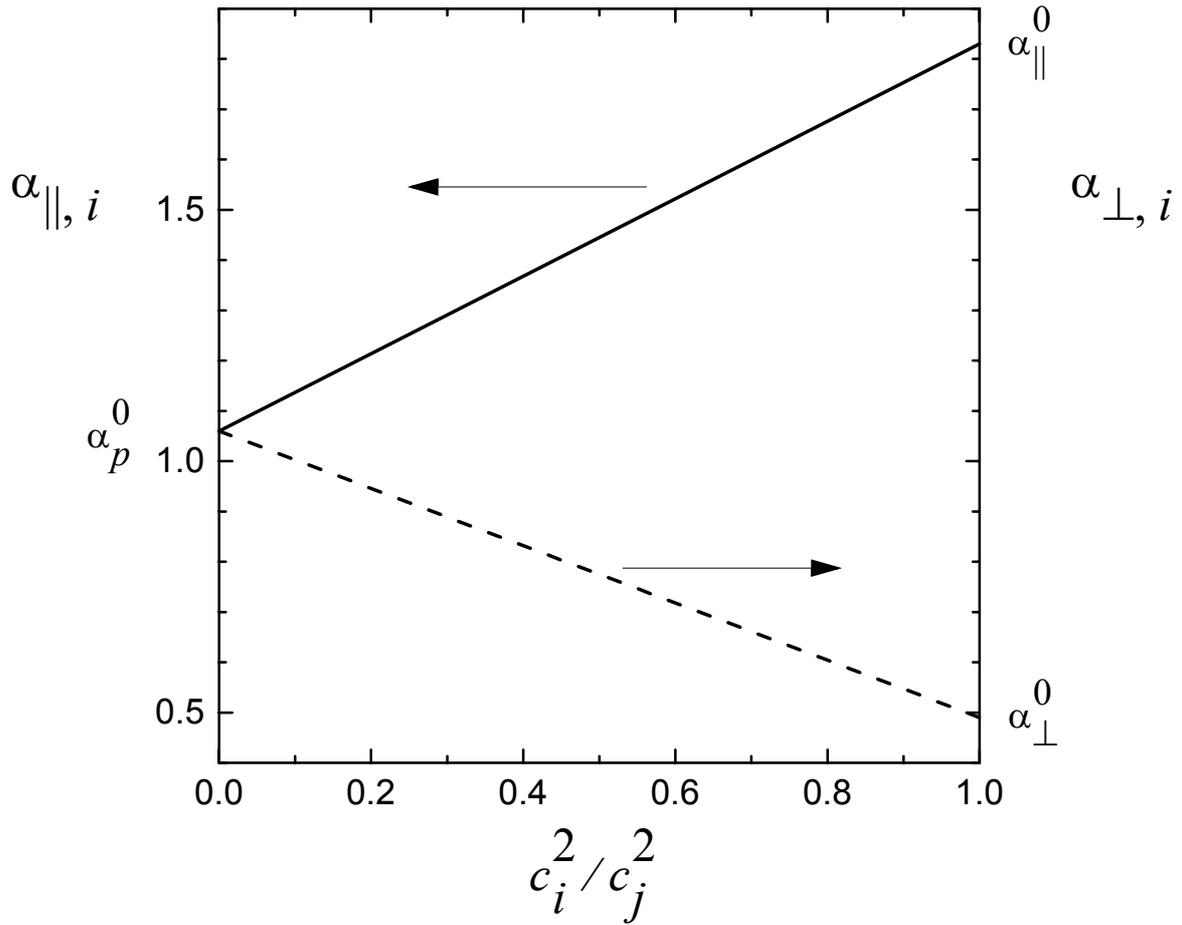

**Figure 4.** When an electron cloud is delocalized over two atoms $i$ and $j$, on which we find different electron densities, the normalized polarizability components parallel ($\alpha_{\parallel,i}$) and perpendicular ($\alpha_{\perp,i}$) to the long axis of the cloud are obtained by linear interpolation between a "normal" $\pi$ bond and a single $p$ orbital, according to the ratio of the electron densities at the two carbons (see text).

Next we multiply these "normalized" polarizabilities, $\alpha_{\parallel}$ and $\alpha_{\perp}$, with the electron densities at the respective carbons (the polarizability is proportional to the polarizable charge). Thus, we obtain for the longitudinal components of the polarizabilities at the atoms $i$ and $j$:

$$\alpha_{\parallel,i} = c_i^2 \left( \alpha_p^0 + (\alpha_\parallel^0 - \alpha_p^0) \frac{c_i^2}{c_j^2} \right) \tag{10}$$

$$\alpha_{\parallel,j} = c_j^2 \left( \alpha_p^0 + (\alpha_\parallel^0 - \alpha_p^0) \frac{c_i^2}{c_j^2} \right) \tag{11}$$

*i* and *j* must be chosen such that $c_i^2 < c_j^2$. The vertical components are given similarly:

$$\alpha_{\perp, i} = c_i^2 \left( \alpha_p^0 + (\alpha_\perp^0 - \alpha_p^0) \frac{c_i^2}{c_j^2} \right) \tag{12}$$

$$\alpha_{\perp, j} = c_j^2 \left( \alpha_p^0 + (\alpha_\perp^0 - \alpha_p^0) \frac{c_i^2}{c_j^2} \right) \tag{13}$$

The total atomic polarizability of atom *i* is a second-rank tensor. For a polarizability tensor of cylindrical symmetry, the effective direction-dependent scalar polarizability can, therefore, be approximated as[26]

$$\alpha_i = \alpha_{\parallel, i} \cos^2 \varphi + \alpha_{\perp, i} \sin^2 \varphi . \tag{14}$$

where $\varphi$ is the angle between the long axis of the electron cloud and the direction vector connecting the carbon atom of that cloud and the solvent unit. Explorative calculations with an elliptical model for the polarizabilities yielded essentially identical values.

For an electron cloud centered on only one atom, as for perylene atoms 1, 8, 9, and 16 in the LUMO (see Figure 2 and Figure 3), a singly occupied *p* orbital with an isotropic polarizability is used:

$$\alpha_i = \alpha_p^0 c_i^2 . \tag{15}$$

In summary, Method A consists of first calculating the two principal polarizability components or the isotropic polarizability for each atom of perylene, for both, HOMO and LUMO, using equations 10-13 and eq 15. The total spectral shift arising from all solvent units *B* is then given by eq 9, where the summation runs over all perylene carbons *i*. Carbon atoms 17 to 20 on the long axis (see Figure 2) can be omitted, since they lie in a nodal plane. The polarizabilities $\alpha_i^{excited}$ and $\alpha_i^{ground}$ for perylene atom *i* with respect to the solvent unit considered are calculated from eq 14 and eq 15.

**Table 4:** Experimental and computational spectral shifts (in cm$^{-1}$) for well defined inclusion sites of perylene in solid *n*-alkanes and in dilute liquid solutions. The computational results were obtained with the methods described in the text and that by Shalev, Ben-Horin, Even, and Jortner (Refs. 3, 8, 9). The accuracy of the method is measured by the root-mean-square deviation of the predicted from the observed shifts. The transition in vacuum occurs at 24'070 cm$^{-1}$.

| system or inclusion site | exp. | method[a] | | | | | |
|---|---|---|---|---|---|---|---|
| | | A | B | C | D | E | SBEJ |
| k [cm$^{-1}$/Å$^2$] | – | 29265 | 24229 | 7367 | 10599 | 23338 | 0.3094 |
| *inclusion sites in the solid* | | | | | | | |
| site 1 hexane[b] | *-1657* | *-1657* | *-1657* | *-1657* | *-1657* | *−1657* | *−1657* |
| site 2 hexane | -1596 | -1593 | -1582 | -1597 | -1581 | −1644 | −1537 |
| site 3 hexane | -1532 | -1532 | -1563 | −1596 | -1558 | −1523 | −1527 |
| site 1 heptane | -1605 | -1601 | -1619 | −1606 | -1574 | −1560 | −1570 |
| site 1 octane | -1565 | -1570 | -1589 | −1570 | -1536 | −1514 | −1544 |
| site 1 nonane | -1540 | -1558 | -1592 | −1559 | -1512 | −1463 | −1524 |
| rms deviation | - | ±8 | ±30 | ±18 | ±26 | ±51 | ±62 |
| *liquid solutions* | | | | | | | |
| *n*-hexane | -1071 | -1072 | −1073 | −1072 | −1047 | −1063 | −938 |
| *n*-heptane | -1095 | −1131 | −1127 | −1126 | −1108 | −1125 | −997 |
| *n*-octane | -1121 | −1148 | −1148 | −1150 | −1126 | −1144 | −1022 |
| *n*-nonane | -1142 | −1192 | −1192 | −1185 | −1164 | −1183 | −1067 |
| *n*-decane | -1163 | −1203 | −1209 | −1206 | −1183 | −1200 | −1088 |
| rms deviation | - | ±35 | ±36 | ±33 | ±18 | ±30 | ±98 |

[a] The column labels refer to the naming of the method in the text, given in the section headings (A: Section 5.1, B: Section 5.2, C: Section 5.3, D: Section 5.4, E: Section 5.5). The column label "SBEJ" refers to results obtained with the perturbation summation by Shalev, Ben-Horin, Even, and Jortner (Refs. 3, 8, 9). [b] This site is used to determine the factor *k* in the computation, which is the only adjustable parameter.

19 of 25

The method outlined above applies to any structure containing perylene and a nonpolar solvent. We have applied it to two different situations: (1) the Shpol'skii system perylene in solid *n*-alkanes, and (2) dilute solutions of perylene in *n*-alkanes. Computation results for some prominent inclusion sites of perylene in solid *n*-alkane matrices[8] and in solution at standard temperature and pressure are displayed in Table 4 together with the experimental values. The agreement with the measurement data is very good.

In the following sections, the sensitivity of the method towards greater simplifications is examined.

### 5.2. Method B: Atom Centered Polarizabilities Only

Consider the spatial extension and the shape of the electron distribution, but only take account of the local electron densities centered at each atom of the chromophore, summing over explicit two-atom interactions only in eq 9. The polarizability difference of each carbon atom of the aromatic molecule in the two electronic states is thought to be proportional to that atom's electron density only. One can then use the electron density itself, e. g., that for the HOMO, in place of $\alpha_A^{excited} - \alpha_A^{ground}$ because we are using a fitting factor $k$ to match the calculated shift for the inclusion site 1 in *n*-hexane to experiment. In this case the formula for the shift calculation for molecule $A$ and all $n$ solvent units $B$ simplifies considerably:

$$\delta E = -k \sum_n \alpha_{B_n} \sum_i \frac{c_i^2}{r_{A_i, B_n}^4} \tag{16}$$

Despite its simplicity, this method gives excellent results, of a precision not significantly lower than those from the perturbation treatments,[8, 9] as can be seen in Table 4.

### 5.3. Method C: Atom Centered Polarizabilities With an Empirical Geometry Factor

A cumulative polarizability factor might improve Method B. It is based on the following considerations: a solvent unit positioned above the plane of the aromatic molecule will interact with fewer polarizable electron clouds in the chromophore than if it were located in the plane of the aromatic molecule. Trying to account for this effect, we suggest an empirical geometric factor, $2 - |\cos\gamma_i|$, where $\gamma_i$ is the angle between the normal to the plane of the perylene molecule and the distance vector between perylene carbon $i$ and the solvent unit. This factor is equal to 1 if the connector between atom *i* and solvent unit is perpendicular to the plane and equal to 2 if the solvent atom lies in the perylene plane, and

$$\delta E = -k \sum_n \alpha_{B_n} \sum_i \frac{c_i^2(2 - |\cos\gamma_i|)}{r_{A_i, B_n}^4} \qquad (17)$$

This formula gives results with somewhat improved precision compared to Method B (see Table 4).

### 5.4. Method D: Four-center Model with Interaction on Both Sides of the Plane

A further simplification begins with Model B (eq 16), but takes into account only the four symmetrically distinct carbon atoms of perylene that have the highest electron density. These form the outer carbons, numbered 1, 8, 9, and 16 (see Figures 2 and 3). Together they account for 40 % of the total electron density and are arranged symmetrically around the central aromatic ring. This simplification is rather crude, but the values obtained are still of reasonable quality (see Table 4).

An attempt to refine this model was made taking into account that the *p* electron clouds of the chromophore are divided into two enantiomeric parts, below and above the molecular plane, which have equal polarizabilities but contribute differently to the dispersion energy because their distance to the particular solvent unit is different. The two parts are displaced by ca. ±60 pm from atom *i* in the perylene plane, a little less than half the aromatic C–C bond length of 140 pm. This refinement was found not to cause significant changes in the results.

### 5.5. Method E: Symmetric Electron Clouds

The final simplification involves the assumption that every electron cloud that is distributed over two carbon atoms be symmetric with respect to the midpoint between these two atoms (as if the orbital coefficients at both atoms were the same). This is equivalent to modeling the electron distribution with ordinary $\pi$ bonds with lower overall electron density. Furthermore, these two-center clouds are represented by a single polarizable entity in the middle of the bond. Equations 10 to 13 simplify to

$$\alpha_{\|,m} = \alpha_\|^0 (c_i^2 + c_j^2) \qquad (18)$$

$$\alpha_{\perp,m} = \alpha_{\perp}^{0}(c_i^2 + c_j^2) \tag{19}$$

where $i$ and $j$ denote the two atoms of the cloud. The summation in eq 9 involves fewer terms since only the mid-points of C-C bonds are considered. For the single atom clouds in the LUMO, eq 15 still applies.

Inspecting the squares of the orbital coefficients at perylene carbon atoms 1 and 2 in the HOMO (they differ by about one order of magnitude, see Table 2) indicates that this is a severe approximation; indeed, the results thus obtained deviate often more than 50 cm$^{-1}$ from the experimental results (see Table 4).

## 6. Conclusions

We have presented a new, accurate approach to the calculation of solvatochromic UV/Vis spectral shifts of an aromatic molecule in a nonpolar solvent matrix. In the case of perylene, the $\pi \rightarrow \pi^*$ transition can be reduced to HOMO and LUMO and the solely geometry-based description with Huckel's theory leads to a precise reproduction of numerous spectral shifts. One fit parameter is contained, used to reproduce one of the well-understood shifts of that chromophore. As a general feature, it is striking that only the connectivity of the aromatic carbons enters the estimate of the local electron densities on each aromatic carbon. We take account of the natural shape of the electron clouds given by the sign of the orbital coefficients, which is missing in the alternative approaches by perturbation theory. The agreement with the experimental results is very good, and our method is computationally much less demanding than previous alternatives. The level of simplifications allow one to choose between accuracy and simplicity.

The second achievement of our spectral shift analyses is that the dispersive van-der-Waals energy falls off with a distance dependence of $1/r^4$ in the distance regime of 300 to 1000 pm, which is most important for modeling intermolecular interactions. The commonplace $1/r^6$ dependence was tested as well and found not to be appropiate. Hirschfelder, Curtiss, and Bird (p 27/28 in ref. 26) stated that the energy between two permanent electric dipoles at shorter distances, where their relative orientation is important, may fall off as $1/r^3$, and only approaches $1/r^6$ at longer distances. The most influential range of our dispersive intermolecular interactions is between 300 pm and 1000 pm so that (dipole) orientation effects are essential, since atomic radii or bond lengths have a magnitude around 100 to 200 pm. We may tentatively explain our detected $1/r^4$ distance dependence by the presence of weak permanent dipoles in the solid and liquid structures. A further investigation of this finding, e. g., for noble gases, and the possibility of more accurate parametrizations for complex force fields seem to be an interesting challenge.

An often used perturbation-theory approach, the SBEJ method[3] (see Table 4), fares only reasonably well in solid matrices, and is computationally very demanding. It seems to be less well-suited for liquid structures, where even our dramatically simplified models, e. g., Model D that only includes 8 terms per solvent unit in the summation, give better values when compared to experiment. It must be said, however, that the system-size sensitivity of our results have not yet been sufficiently explored for liquid systems (in contrast to the crystalline systems considered); a full investigation of this point would, however, require extensive further calculations. The advantage of our methods lie in their theoretical simplicity, ease of implementation in a computational program for a particular system, and efficiency.

Moreover, it would be interesting to expand these ideas to more complicated systems, e. g., those with polar constituents or hydrogen bonding.

## Acknowledgments

We are grateful about suggestions from Prof. H. C. Oettinger, Department of Materials, ETH Zurich, Prof. A. Nikolaidis, Department of Chemistry, University of Cyprus, Nicosia, and Alois Renn, Department of Chemistry, ETH Zurich. We also acknowledge the ETH Office of Student Mobility, the Swiss National Science Foundation (Schweizerischer Nationalfonds), and the German National Merit Foundation (Studienstiftung des Deutschen Volkes) for support.